\pdfoutput=1

\documentclass{jfm}
\usepackage{graphicx}
\usepackage{epstopdf, epsfig}

\usepackage{graphicx}
\usepackage{subfigure}
\usepackage{amssymb}
\usepackage{amsmath}
\usepackage{mathrsfs}
\usepackage{placeins}
\usepackage{natbib}
\usepackage{caption3}
\bibpunct{(}{)}{,}{s}{,}{,}

\usepackage[bookmarksnumbered=true,breaklinks=true,colorlinks,citecolor=blue,linkcolor=blue]{hyperref}
\usepackage[geometry]{ifsym}
\makeatletter

\newcommand{\numtoRoman}[1]{\expandafter\@slowromancap\romannumeral #1@}

\renewcommand{\figurename}{Figure}
\newcommand\solidrule[1][8mm]{\rule[0.5ex]{#1}{.4pt}}

\usepackage[usenames]{color}

\makeatother

\shorttitle{Effects of slip surfaces on turbulent transition}
\shortauthor{E. A. Davis and J. S. Park}

\title{Dynamics of laminar and transitional flows over slip surfaces: effects on the laminar-turbulent separatrix}

\author{Ethan A. Davis
  \and Jae Sung Park
  \corresp{\email{jaesung.park@unl.edu}}}

\affiliation{Department of Mechanical and Materials Engineering, University of Nebraska-Lincoln,
Lincoln, NE 68588-0526, USA}

\begin{document}

\maketitle

\begin{abstract}
The effect of slip surfaces on the laminar-turbulent separatrix of plane Poiseuille flow is studied by direct numerical simulation. In laminar flows, the inclusion of the slip surfaces results in a drag reduction of over 10\%, which is in good agreement with previous studies and the theory of laminar slip flows. Turbulence lifetimes, the likelihood that turbulence is sustained, is investigated for transitional flows with various slip lengths. We show that slip surfaces decrease the likelihood of sustained turbulence compared to the no-slip case, and the likelihood is further decreased as slip length is increased. A more deterministic analysis of the effects of slip surfaces on a transition to turbulence is performed by using nonlinear travelling wave solutions to the Navier-Stokes equations, also known as exact coherent solutions. Two solution families, dubbed P3 and P4, are used since their lower-branch solutions are embedded on the boundary of the basin of attraction of laminar and turbulent flows \citep{Park2015jfm}. Additionally, they exhibit distinct flow structures -- the P3 and P4 are denoted as core mode and critical layer mode, respectively. Distinct effects of slip surfaces on the solutions are observed by the skin friction evolution, linear growth rate, and phase-space projection of transitional trajectories. The slip surface appears to modify transition dynamics very little for the core mode, but quite considerably for the critical layer mode. Most importantly, the slip surface promotes different transition dynamics -- an early and bypass-like transition for the core mode and a delayed and H- or K-type-like transition for the critical layer mode. We explain these distinct transition dynamics based on spatiotemporal and quadrant analyses. It is found that slip surfaces promote the prevalence of strong wall-toward motions (sweep-like events) near vortex cores close to the channel centre, inducing an early transition, while long sustained ejection events are present in the region of the $\Lambda$-shaped vortex cores close to the critical layer, resulting in a delayed transition. This should motivate flow control strategies to fully exploit these distinct transition dynamics for transition to turbulence.
 
\end{abstract}

\begin{keywords}
%Authors should not enter keywords on the manuscript, as these must be chosen by the author during the online submission process and will then be added during the typesetting process (see http://journals.cambridge.org/data/\linebreak[3]relatedlink/jfm-\linebreak[3]keywords.pdf for the full list)
\end{keywords}

\section{Introduction}
In the last few decades, the allure of slip surfaces for use in a multitude of flow applications has grown \citep{Golovin2016b}. Of particular interest is the demonstrated ability of slip surfaces in reducing frictional resistance, or drag, in many different configurations for both laminar and turbulent flows. Many studies have achieved significant drag reduction via slip by placing hydrophobic surfaces at the walls \citep{Luchini1991, Watanabe1996, Min2004b, Ou2004, BrinkerPRL2006, MaynesPoF2006, Bocquet2011b, Seo2018d}. However, a smaller number of studies have been performed on the effect of slip surfaces on the transition to turbulence with most limited to linear stability analysis of the linearized Navier-Stokes equations. Two-dimensional \citep{Spille2001, Yu2016} and three-dimensional \citep{Min2005c, Lauga2005, You2007} analyses on the stability of flows have been carried out. While most of these studies have shown a stabilizing effect of slip surfaces leading to a delayed transition, others have found that these surfaces can induce or amplify instabilities and, subsequently, trigger early transition \citep{Chu2004, Chai2019}. Studies on a bluff body with a superhydrophobic coating have shown different results on the flow separation -- delayed separation \citep{Gruncell2013} and promoted separation \citep{Castagna2018}. Furthermore, a recent study by \citet{Picella2019} found that slip surfaces may be effective at delaying transition for near-wall perturbations similar to ones observed in H-/K-type transition scenarios \citep{Klebanoff1962} but are rendered ineffectual for free-stream, or non-modal, perturbations similar to ones observed in a bypass transition scenario \citep{Morkovin1985}. As such, there exists no strict consensus on the effects of slip surfaces on the transition to turbulence, or in which particular transition scenarios slip surfaces may be considered effective.

Interestingly, there have been doubts on the validity of the no-slip boundary conditions at the microscopic scale, showing that under certain circumstances, fluids can slip against solid surfaces \citep{Granick2003, Squires2005}. Various factors including wettability, surface roughness, presence of gaseous layers, and impurities can facilitate a non-zero fluid velocity at solid surfaces \citep{deGenne2002, Quere2005, Neto2005}. The idea of slip was first introduced by Navier, quantifying the slip velocity by the idea of a slip length \citep{Navier1823}. The slip length relates the velocity of the fluid at the wall to the wall shear rate as
\begin{equation} \label{eq:slip}
u_s= b\left.\frac{\partial{u}}{\partial{y}} \right|_w,
\end{equation}
where $b$ is the slip length or Navier’s slip coefficient. The slip at the solid surface tends to increase the velocity at the wall, leading to skin-friction reduction. \cite{Min2005c} investigated temporal transition to turbulence, where the linear stability analysis was applied to the linearized Navier-Stokes equations for a two-dimensional case, showing that the critical Reynolds number increases with the streamwise slip length and decreases with spanwise slip length. A combination of the two results in a critical Reynolds number between the no-slip and limiting streamwise cases. More recently, superhydrophobic surfaces, which are a combination of surface chemistry and surface roughness at micro- and/or nano-scales, have been introduced for drag reduction \citep{RothsteinARFM2010}. These surfaces appear to lower the free energy of an air-water interface, producing a very high contact angle at the surface. There have been, of course, many studies on the effects of superhydrophobic surfaces on skin-friction reduction in laminar and turbulent flows \citep{Park2014, Lee2015, Jung2016, Lee2016e, Ling2016, Gose2018, Seo2018d, Fairhall2019}. Thorough reviews of the effect of slip and superhydrophobic drag reduction on laminar and turbulent flows are given by \citet{RothsteinARFM2010}, \citet{Abdulbari2013b}, and \citet{Lee2016e}. It is worth noting that there is a recent study that performs direct numerical simulations of a turbulent channel flow to predict the effective slip length and drag reduction with a lubricated micro-groove surface \citep{Chang2019}.

In addition to laminar and turbulent flows, a classical problem in wall-bounded shear flows is the transition to turbulence \citep{Avila2011, Barkley2016, Sano2016}. The fundamental question of what causes a flow to undergo a change from a highly ordered laminar state in space and time to a highly disordered turbulent state in space and time has been puzzled over since Osborne Reynolds in 1883 \citep{Reynolds1883}. There have been several approaches to explore the nature of transition. The directed percolation has been proposed to conjecture the spatiotemporal intermittency observed in a transitional flow \citep{Pomeau1986, Sipos2011, Allhoff2012, Lemoult2016}. The theoretical and experimental studies based on puffs and slugs have also been explored to study the dynamics of transition to turbulence \citep{Wygnanski1973, Nishi2008, Barkley2011, Barkley2015, Shih2016}. Recently, the dynamical systems idea has emerged, enabling computation of non-trivial invariant solutions to the Navier-Stokes equations \citep{Kawahara2012}. These solutions, considered as the building blocks of turbulence, have advanced our understanding of chaotic spatiotemporal flows \citep{Suri2017, Tithof2017, Park2015jfm, Park2018}. These spatially and temporally well-organized solutions have been observed in all turbulent shear flows, both in experiments and simulations \citep{Hof2004, Gibson2009}, and have been used to help provide qualitative and quantitative information about the transition to turbulence.

We focus here on the dynamical systems approach, which has greatly advanced the understanding of the nature of the transition to turbulence \citep{Kerswell2005, Eckhardt2007, Lustro2019}. Specifically, the discovery of three-dimensional fully nonlinear traveling wave (TW) solutions to the Navier-Stokes equations has enabled \textit{a prior} study of the transition to turbulence. These solutions are also denoted as exact coherent states (ECS) since they capture the essential structural and statistical features of a turbulent flow, featuring the near-wall self-sustaining process \citep{Waleffe2001}. They primarily arise in pairs from a saddle-node bifurcation at a particular Reynolds number, consisting of upper- and lower-branch solutions -- both of which are unstable. These lower- and upper-branch TW solutions can, thus, be thought of as saddle points in the state-space and turbulent trajectory moves dynamically between these saddle points \citep{Park2015jfm}. More interestingly, the onset Reynolds number for TW solutions to come into existence is quantitatively in good agreement with the Reynolds number for transition to turbulence, corresponding to a spanwise period of 100 wall units \citep{Carlson1982}. Interestingly, the subcritical nature of these bifurcations of TW solutions has shed light on some features of subcritical turbulent transition such as patterned turbulence \citep{Tuckerman2014}. Most of these ECSs have been found in so-called minimal flow units –-- they are spatially periodic in the unbounded dimensions of the domain with periods that roughly correspond to the smallest length scales at which turbulence can persist \citep{Jimenez1991}. In addition, localized ECSs in extended domains have also been found to show flow structures resembling the observed spatiotemporal intermittency and the laminar-turbulent patterning that are a common feature of turbulence near transition \citep{Avila2013, Brand2014}. \cite{Avila2013} found a particular family of localized ECSs for pipe flow that shows features similar to puffs, which resemble localized turbulent regions in a transitional pipe flow.  \cite{Zammert2014} also found a family of localized ECSs for a plane Poiseuille flow. However, their connections to turbulence transition are still unclear.

Another important issue regarding ECSs is their connection to the laminar-turbulent boundary, or separatrix. This boundary separates the basins of attraction of laminar and turbulent flows \citep{Schneider2007, Duguet2008}, and this boundary is inherently unstable. Initial conditions on the turbulent side of the boundary become turbulent, while those on the laminar side laminarize. Initial conditions on this boundary stay on it -- they neither become turbulent nor do they relaminarize.There are theoretical arguments that the stable manifold of the lower-branch ECS forms a part of this boundary \citep{Wang2007, Kawahara2005}. If an ECS on the basin boundary has a single unstable eigenvalue and many stable eigenvalues, there are many ways in which trajectories can approach it, but there is only one way to leave. Such ECS has been called an edge state \citep{Skufca2006}. Some of the lower-branch ECSs are also found to lie on the basin boundary, but they have multiple unstable eigenvalues so are not edge states \citep{Park2015jfm}. The dynamics of trajectories on or near the basin boundary are intermediate between laminar and turbulent flows and thus thought to play an important role in the dynamics of transition to turbulence.

Because of the clear importance of the lower-branch ECSs for the transition to turbulence and proximity to the basin boundary, the present work focuses on the effects of slip surfaces on these ECSs embedded on the laminar-turbulence boundary. It will provide an excellent simplified, yet still exact, model flow for studying the nature of the transition to turbulence. The leading order effect of the slip surfaces on the ECS is one important focus of the present study for identifying dynamics of the transition. The connection between ECSs and slip surfaces has yet to be fully explored and will be investigated in the present study. The nature of turbulence transition will be identified with respect to ECSs on the laminar-turbulent separatrix found by \cite{Park2015jfm}.

We focus here on a channel flow of an incompressible Newtonian fluid with dynamic viscosity $\mu$, density $\rho$ and kinematic viscosity $\nu = \mu / \rho$ in a channel of half-height $h$. In this geometry, the critical Reynolds numbers are $Re = U_c h/\nu \approx 1000$ and $Re_{\tau} = u_{\tau}h/\nu \approx 45$ based on the laminar centerline velocity $U_c$ and friction velocity $u_{\tau}$, respectively \citep{Carlson1982, Pope2000}. The laminar and transition flow regimes in a range of $120 < Re < 1800~(8 < Re_{\tau} < 85)$ will be considered in the present study. Two lower-branch ECSs (labelled P3 and P4 below) will be considered as they are on the basin boundary -- P3 is indeed an edge state as it has only one unstable eigenvalue. It is worth noting that, to our knowledge, there have been only two modes of the exact coherent solutions \citep{Waleffe2001, Gibson2009, Viswanath2009, Nagata2013, Gibson2014, Park2015jfm}, namely core mode and critical layer mode. P3 and P4 solutions display characteristics of each mode, respectively. Therefore, by examining the P3 and P4 solutions, it is believed that the essential effects of slip surfaces on transition dynamics will be encapsulated.

This paper is organized as follows: Section \ref{sec:formulation} presents the problem formulation for the current study. A validation of the current simulations is given in Section \ref{sec:laminar} by comparing drag reduction of laminar flows. To better understand the effect of the slip on the transition, the effect of slip on turbulence lifetimes is presented in Section \ref{sec:lifetime}. Additionally, the effect of slip surfaces on the transition to turbulence is studied using traveling wave solutions to the Navier-Stokes equations in Section \ref{sec:tws}. Finally, a summary and implications of the present investigation are given in Section \ref{sec:conclusion}.
 
\section{Problem Formulation}\label{sec:formulation}
We consider an incompressible Newtonian fluid in the plane Poiseuille (channel) geometry, driven by a constant volumetric flux $Q$. The $x$, $y$, and $z$ coordinates are aligned with the streamwise, wall-normal, and spanwise directions, respectively. Periodic boundary conditions are imposed in the $x$ and $z$ directions with fundamental periods $L_x$ and $L_z$, and streamwise Navier slip conditions are imposed at the walls $y=\pm h$, where $h=L_y/2$ is the half-channel height. The laminar centerline velocity for a given volumetric flux is given as $U_c = (3/4)Q/h$. Using the half-height $h$ of the channel and the laminar centerline velocity $U_c$ as the characteristic length and velocity scales, respectively, the nondimensionalized Navier-Stokes equations are then given as
\begin{equation}
\nabla \cdot \boldsymbol{u}=0,\quad
\frac{\partial \boldsymbol{u}}{\partial t}+\boldsymbol{u}\cdot\nabla \boldsymbol{u}=-\nabla p+\frac{1}{Re_c}\nabla^2\boldsymbol{u}.
\end{equation}
Here, we define the Reynolds number for the given laminar centerline velocity as $Re_c = U_c h/\nu$, where $\nu$ is the kinematic viscosity of the fluid. Characteristic inner scales are the friction velocity $u_{\tau}=(\bar{\tau}_w/\rho)^{1/2}$ and the near-wall length scale or wall unit $\delta_{\nu} = \nu/u_{\tau}$, where $\rho$ is the fluid density and $\bar{\tau}_w$ is the time- and area-averaged wall shear stress. As usual, quantities nondimensionalized by these inner scales are denoted with a superscript ``+". The friction Reynolds number is then defined as $Re_{\tau}=u_{\tau}h/\nu=h/\delta_{\nu}$. Streamwise Navier slip conditions are prescribed as equation (\ref{eq:slip}) at both top and bottom walls by an effective homogeneous slip length, $L_s = b/h$. To verify that the slip length can be realistically obtained by practical slip surfaces with roughness features, the length scale $L^+$ of the micro roughness can be calculated using equation (2.4) of \citet{Picella2019}, enabling a direct comparison with ones in the literature. Using the largest slip length studied ($L_s = 0.02$) and the solid fraction $\phi_s = 0.25$ used in \citet{Min2005c}, the largest texture size of the current study is $L^+ \approx 7$, which ensures that the homogeneous slip surface employed in the present study would provide virtually the same outcomes resulting from employing a heterogeneous microtextured slip surface \citep{Ybert2007, Seo2018, Picella2019}.

Simulations are performed using the open source code \textit{ChannelFlow} written and maintained by \cite{channelflow}. In this study, we focus on the domains of $L_x \times L_y \times L_z = 2\pi \times 2 \times \pi$ and $L_x \times L_y \times L_z = \pi \times 2 \times \pi/2$, which are the same box sizes as the TW solution families dubbed P3 and P4, respectively \citep{Park2015jfm}. A numerical grid system is generated on $N_x \times N_y \times N_z$ (in $x$, $y$, and $z$) meshes, where a Fourier-Chebyshev-Fourier spectral spatial discretization is applied to all variables. A typical resolution used is $(N_x, N_y, N_z) = (48, 81, 48)$. The numerical grid spacing in the streamwise and spanwise direction are $\Delta x_{min}^{+} \approx 8.2 (4.7)$, $\Delta z_{min}^{+} \approx 4.1  (2.3)$ for the P3 and (P4) cases. The nonuniform Chebyshev spacing used in the wall-normal direction results in $\Delta y_{min}^{+} \approx 0.05$ at the wall and $\Delta y_{max}^{+} \approx 2.5$ at the channel centre for both P3 and P4 cases. For simulations, a range of $120 < Re < 1800~(8 < Re_{\tau} < 85)$ is considered to cover laminar and transition flow regimes. For exact coherent solution, $Re_\tau = 62.52$ and $71.72$ are considered for P3 and P4, respectively, as these are the inherent Reynolds numbers at which the P3 and P4 solutions emerge.

\section{Results and Discussion}\label{sec:results}
\subsection{Laminar drag reduction: a validation}\label{sec:laminar}
\begin{figure}
 \begin{center}\includegraphics[trim= 1 2 4 1, clip, scale=.65]{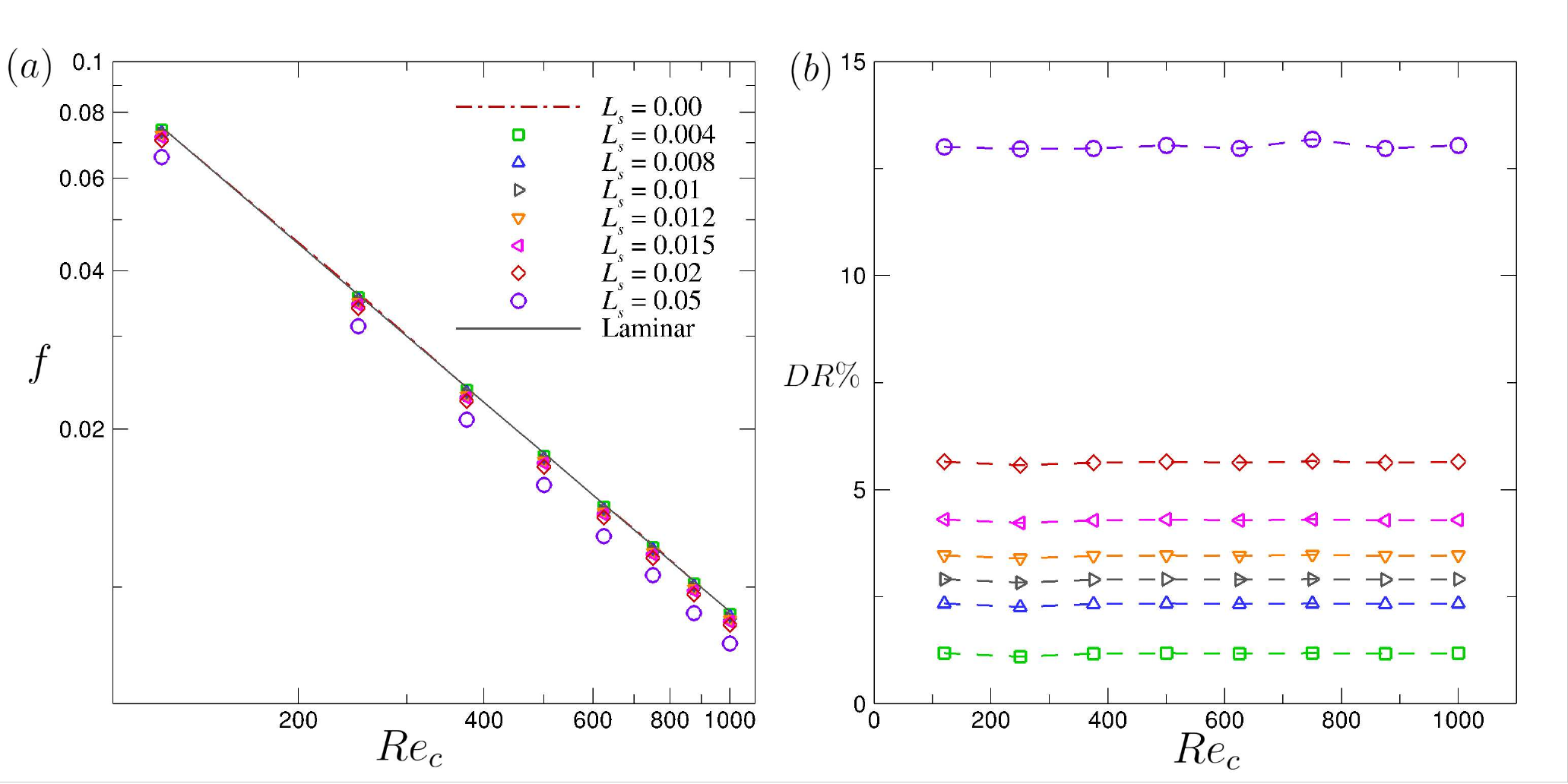}
  \caption{(\textit{a}) Friction factor and  (\textit{b}) drag reduction percentage as a function of Reynolds number for various $L_s$ values. In $(a)$, laminar flow (solid line) corresponds to $f = 9/\Rey$ \citep{Pope2000}. In $(b)$, the dashed lines are shown for readability.
    \label{fig:dr}}
 \end{center}
\end{figure}

For the sake of testing the code, the effect of the slip surfaces on laminar flows was investigated. Drag reduction percentage ($DR\%$) was calculated to compare to previous studies and is given by:

\begin{equation} \label{eq:dr}
DR\%=\frac{f_{0}-f}{f_{0}}\times100,
\end{equation}

where $f$ is the friction factor for the slip surface and $f_{0}$ is the friction factor for the no-slip surface at the same Reynolds number: $f = \bar{\tau}_w/(1/2\rho U_b^2)$, where $U_b$ is the bulk velocity. Figures \ref{fig:dr}$(a)$ and $(b)$ show the friction factor and drag reduction percentage as a function of Reynolds number for various slip lengths, respectively. As seen in figure \ref{fig:dr}$(a)$, the friction factor is shifted downward when slip length is increased, while maintaining the same slope as laminar case. Therefore, the relative change in wall shear stress or pressure drop with Reynolds number remains constant and the drag reduction from equation (\ref{eq:dr}) is constant for each slip length. Figure \ref{fig:dr}$(b)$ confirms that $DR\%$ is almost constant at each slip length regardless of Reynolds number. The same trend of constant drag reduction percentage in the laminar regime has also been observed previously \citep{Ou2004}, where an increase in drag reduction resulted from increasing slip length, alone, and was not a function of flow rate.

\begin{figure}
 \begin{center}
 \includegraphics[trim=2 2 4 4,clip,scale=.55]{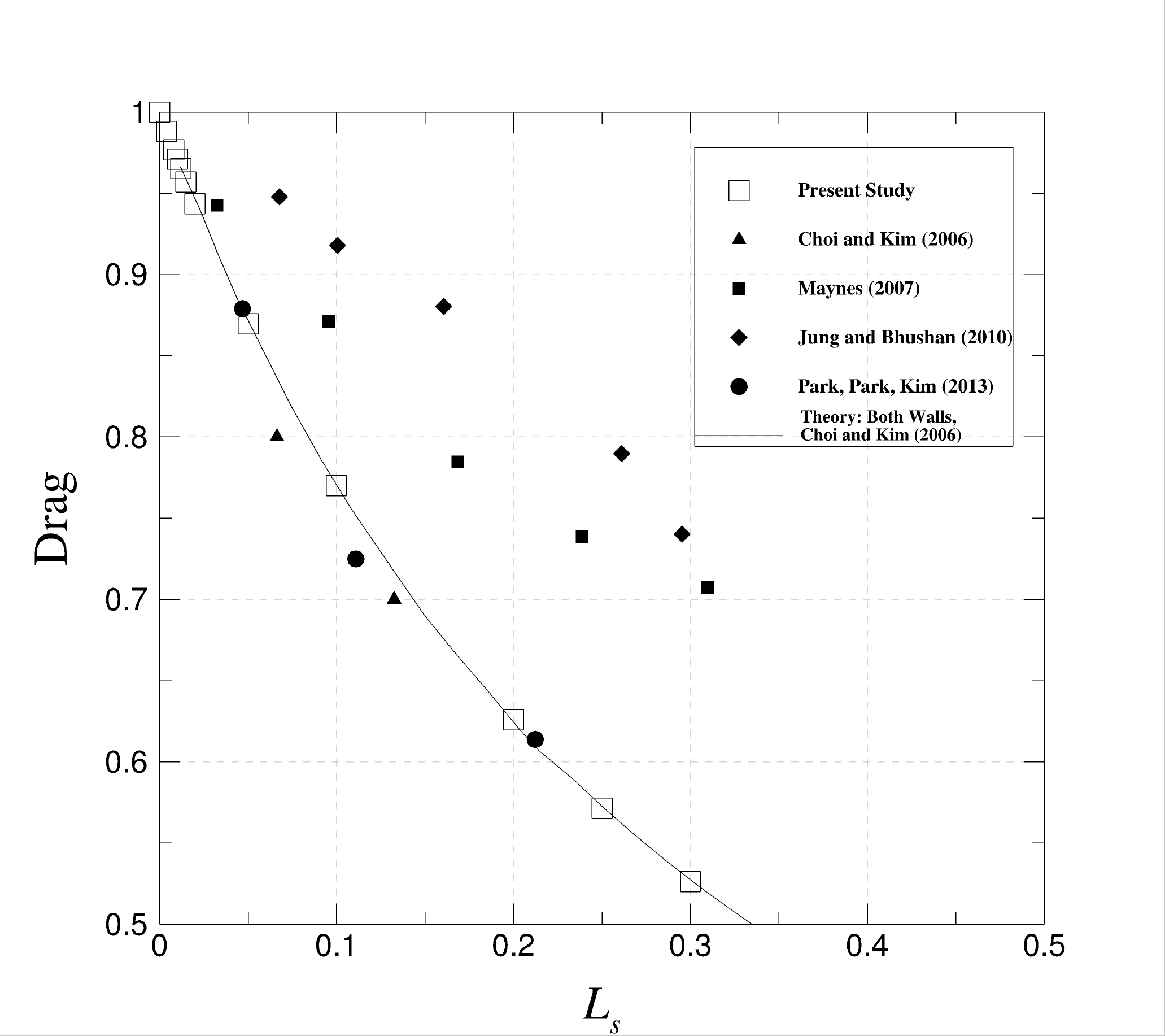}
  \caption{Drag ($f/f_0$) corresponding to the friction factor ($f$) normalized by that of the no-slip surface ($f_0$) as a function of slip length $L_s$ normalized by the channel half-height, $h$: \SmallSquare , present study; \FilledSmallTriangleUp ,  \citep{Choi2006b}; \FilledSmallSquare , \citep{Maynes2007b}; \FilledSmallDiamondshape ,\citep{Jung2010b}; \FilledSmallCircle , \citep{Park2013d}; \solidrule , theory for superhydrophobic surfaces on both walls \citep{Choi2006b}.}
    \label{fig:drag}
 \end{center}
\end{figure}

To further validate the present study, drag values ($f/f_0$) were compared to those of previous studies using superhydrophobic surfaces in laminar flows. Figure \ref{fig:drag} shows the results for the drag value observed in the laminar regime for previous studies (closed symbols) and the current study (open symbols) along with a curve for the theory for the superhydrophobic surfaces on both walls. As expected, the drag values decrease with increasing slip length. Notably, the current study is in great agreement with the theory proposed by \citet{Choi2006b} for superhydrophobic surfaces on both channel walls.

\subsection{Turbulence lifetimes: statistical insight into transition to turbulence over slip surfaces}\label{sec:lifetime}
A common approach used in the study of flow control and its effect on the transition to turbulence is to compute the turbulence lifetime of the flow. This lifetime gives a more physical interpretation of the effects of the flow control on the flow and on the modified distance between the turbulent state and the edge of turbulence in phase space. \citet{Ibrahim2019b} showed that opposition control in both Couette and Poiseuille flows plays a role in increasing the probability of escaping from the chaotic saddle of turbulence in phase space. In this regard, the effect of slip surfaces on altering turbulence lifetime statistics was investigated. Similar to the procedure used by \citet{Ibrahim2019b}, 100 different turbulent flow fields were created by running simulations at an elevated Reynolds number. These 100 flow fields were then used as the initial conditions for the turbulence lifetime study at the nominal Reynolds number (i.e. flow fields created at $Re_c$ = 1900 were used as initial conditions for the turbulence lifetime study at $Re_c$ = 1800). The temporal evolution of the wall shear stress was tracked up to $t$ = 5000 for four different slip lengths: $L_s$ = 0.00, 0.008, 0.01, and 0.02. Note that the slip length for the current study is already in dimensionless form $L_s = b/h$, where $h$ is the channel half-height. If a flow laminarizes, the time it takes for the wall shear stress to become its laminar value is considered the turbulence lifetime. Probability of turbulence can then be computed as a function of time for each value of $L_s$ by defining it as the fraction of the 100 turbulent flow fields that remain turbulent up to a given time $t$. 

\begin{figure}
 \begin{center}\includegraphics[trim= 4 4 4 4, clip, scale=.65]{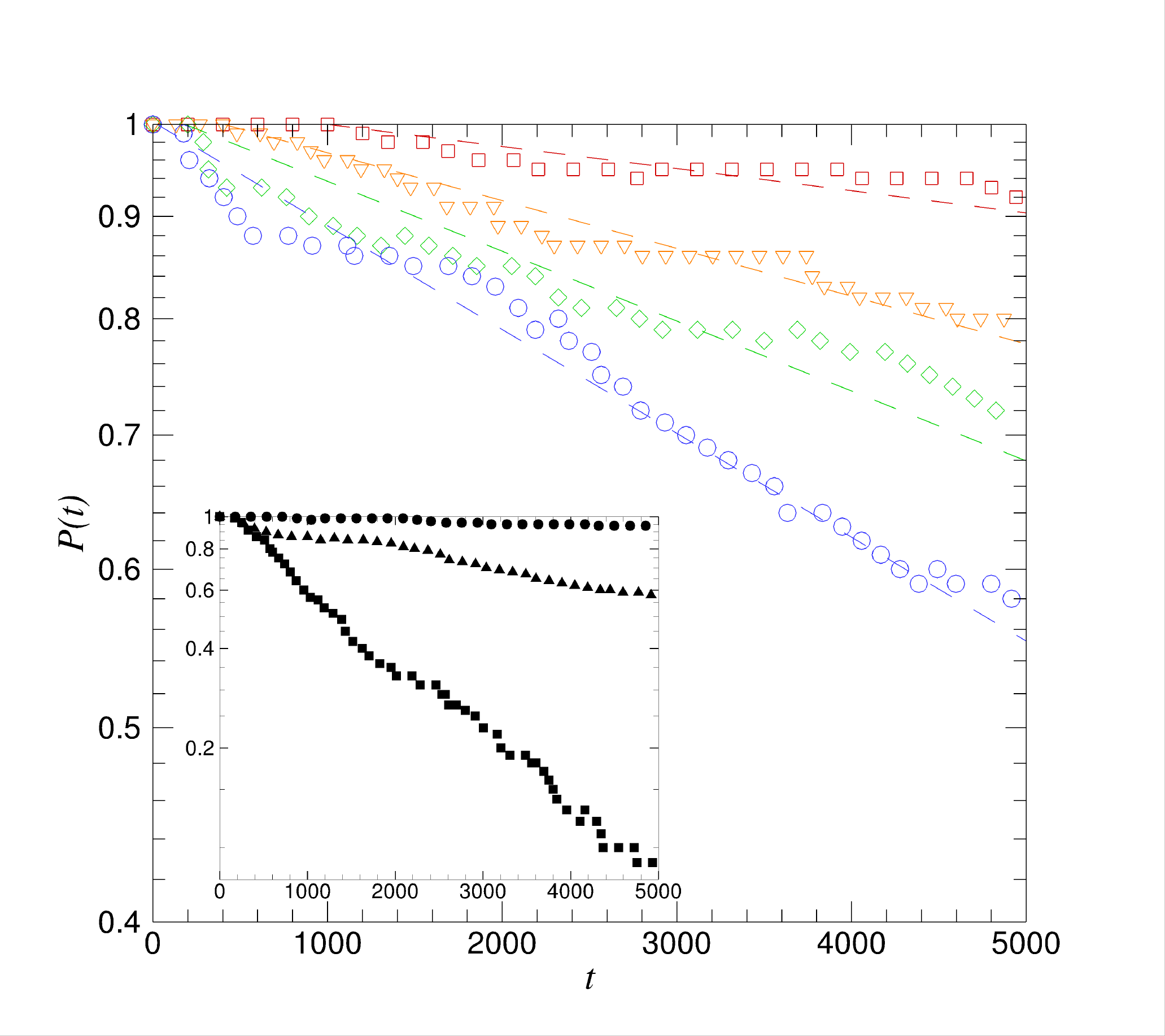}
  \caption{Probability of turbulence lifetime for slip lengths $L_{s}=0.00$ (\SmallSquare), $0.008$ (\SmallTriangleDown), $0.01$ (\SmallDiamondshape), and $0.02$ (\SmallCircle) at $Re_c = 1800$ on semi-logarithmic axes. The dashed lines are shown for readability to help illustrate the exponentially decaying trend. Inset: Dependence of Reynolds number on turbulence lifetime at $Re_c = 1600$ (\FilledSmallSquare), $1800$ (\FilledSmallTriangleUp), and $2000$ (\FilledSmallCircle) at $L_{s} = 0.02$.
    \label{fig:lifetime}}
 \end{center}
\end{figure}

Figure \ref{fig:lifetime} shows the probability of turbulence lifetimes for various Reynolds numbers at slip lengths of $L_{s} = 0.00, 0.01,$ and $0.02$. The probability is found to approximately follow an exponential distribution that is characteristic of turbulence lifetimes \citep{EckhardtARFM2007}. It should be noted that the probability of turbulence does decrease for the no-slip case in the transitional flow regime. Given sufficient time and domain size and the nature of initial conditions, it has been shown that there is a tendency for turbulence to decay as a result of a transient process in shear flows \citep{Schmiegel1997,Bottin1998,Hof2006,Hof2011,Schneider2008}. As intuition might suggest, the probability of sustained turbulence at a given time decreases as slip length is increased, indicating a stabilizing effect on the flow. As was investigated by \citet{Min2005c}, the addition of a streamwise slip velocity stabilizes a flow by which it is likely that the probability of turbulence persisting decreases as streamwise slip length is further increased. While this trend holds for the three Reynolds numbers studied here--$Re_c = 1600, 1800,$ and $2000$--the effectiveness of a given slip surface decreases with increasing Reynolds number. In other words, as Reynolds number increases, larger slip lengths are needed to obtain the same effect. Notably, the slip lengths studied here have little effect on the flow at $Re_c = 2000$, with a final reduced probability of sustained turbulence of $\sim94$\% at the largest slip length examined. Note that the effect of slip surfaces on turbulence lifetime may be interpreted as a reduction of the Reynolds number, which was observed by \citet{Min2004b} and \citet{Fukagata2006}. These results suggest that the phase space of turbulence may be altered by slip surfaces in such a way to facilitate transition back to the laminar state. This same trend in turbulence lifetime was also observed when using increasing levels of opposition control in channel flow \citep{Ibrahim2019b}.
 
\subsection{Exact coherent solutions: a deterministic analysis of the laminar-turbulent separatrix over slip surfaces}\label{sec:tws}

We now discuss the effect of slip surfaces on the laminar-turbulent separatrix using exact coherent solutions or travelling-wave solutions to the Navier-Stokes equations. Two travelling-wave solution families, dubbed P3 and P4, were used as the lower-branch solutions for these two families are embedded in the laminar-turbulent separatrix \citep{Park2015jfm}. The P3 lower-branch solution is indeed an edge state as it has only one unstable eigenvalue in the symmetric subspace, while the P4 lower-branch solution has two unstable eigenvalues in its symmetric subspace. The edge state is particularly important because as defined, it has only one negative eigenvalue, or only one unstable direction, which implies somehow the weakest most marginal structure of the basin boundary. In addition, these two solutions were chosen due to their distinct flow structures -- the P3 and P4 families are denoted as core and critical layer modes, respectively. This implies that the flow structures for the these solutions propagate in either the core, or bulk, of the flow or centered about the critical layer. As aforementioned, to our best knowledge, the exact coherent solutions that have been found thus far have exhibited flow structures that belong to either the core or critical mode. These modes are also analogous to the modal/non-modal perturbations that have been examined in the classical transition scenarios \citep{Wu2009, Picella2019}. Recently, the P4 solution is found to display hairpin-like vortex structures \citep{Shekar2018}. Additionally, a recent study by \citet{Picella2019} found that by using linear stability analysis, slip surfaces have distinct effects on different transition scenarios defined by the initial perturbations applied to the base flow, namely, modal, near-wall perturbations and non-modal, free-stream perturbations. Thus, it is anticipated that the effect of slip surfaces on these solution families exhibit distinct transition dynamics.

Here, we ran simulations using the P3 and P4 lower-branch solutions as initial conditions to investigate the effects of slip surfaces on the separatrix and to elucidate transition dynamics. Using an exact coherent solution (i.e., a traveling-wave solution), a more deterministic approach can be taken in analyzing the effects of slip surfaces. The P3 and P4 lower-branch solutions used are $Re_c$ = 1760 ($Re_\tau = 62.52$) and $Re_c$ = 1800 ($Re_\tau = 71.72$), respectively. These Reynolds numbers were chosen, in particular, because it is close to their bifurcation Reynolds numbers but far enough to ensure that they are embedded within the laminar-turbulent boundary.

\subsubsection{Skin-friction evolution}
\begin{figure}
 \begin{center}\includegraphics[trim=4 4 4 4,clip, scale=.65]{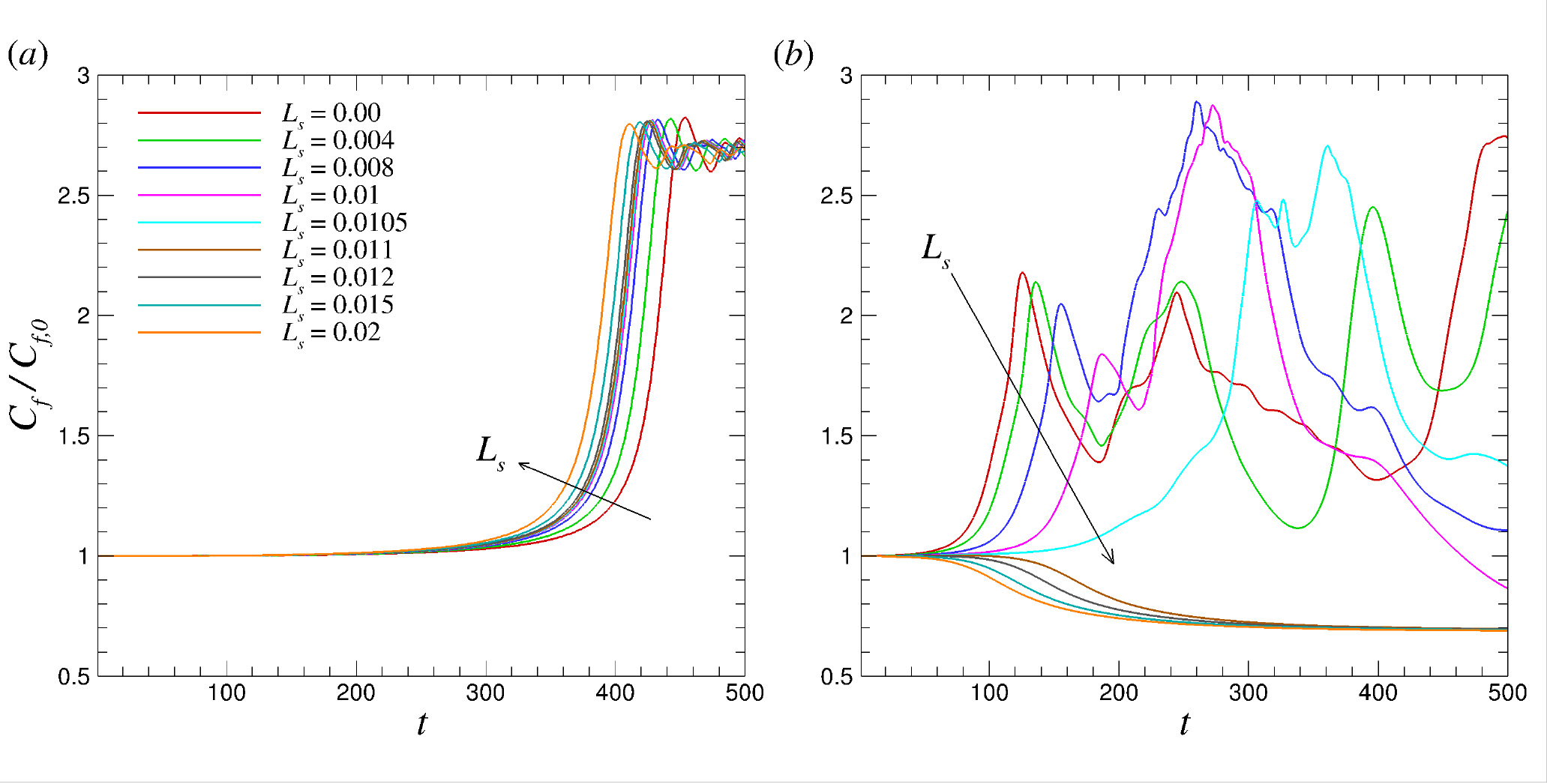}
  \caption{Early-time skin-friction coefficient profiles when using the (\textit{a}) P3  and (\textit{b}) P4 lower-branch travelling-wave solutions as initial conditions. Profiles are normalized by the initial skin-friction coefficient value for each case. The P3 and P4 lower-branch solutions are on the laminar-turbulent separatrix.
    \label{fig:Cf}}
 \end{center}
\end{figure}

Figure \ref{fig:Cf} illustrates the temporal behaviours of the skin-friction coefficient $C_f = \bar{\tau}_w/(1/2\rho U_c^2)$ normalized by initial values of each case using various $L_s$ values. Despite their distinct characteristics, the early-time behavior of the skin friction of the P3 and P4 solutions is similar: (1) an initial stable period, (2) a sharp increase, or a strong turbulent burst, following the stable period, and (3) transition to a fully turbulent flow (for P4, a flow is relaminarized beyond $L_s > 0.0105)$. This temporal behaviour indeed resembles a typical scenario observed in transition to turbulence \citep{Schmid2012}. The strong turbulent burst, as it shall be called presently \citep{Park2018}, is the process of escape out of the exact coherent solution along its most unstable manifold, comprising of the so-called linearly unstable stage followed by the nonlinear evolution stage \citep{Itano01}. The end of this strong turbulent burst is then defined as the time when the skin friction reaches its maximum. It is evident that distinct transition scenarios are observed between the P3 and P4 solutions by the evolution of the skin friction due to inclusion of the slip surfaces. 

Notably, as shown in figure \ref{fig:Cf}, the duration of the initial stable period decreases with slip length for the P3 solution, while the duration of this period increases with slip length for the P4 solution. It suggests that the P3 and P4 solutions experience early transition and delayed transition due to slip surfaces, respectively. For the P3 solution, the slip length appears to have a negligible effect on the reduction in maximum skin friction during the strong turbulent burst as the slip length is further increased. However, the behavior of the strong turbulent burst for the P4 solution is quite different. The maximum skin friction reached during the strong turbulent burst following the initial stable period decreases with increasing slip length. This trend continues until a critical slip length ($L_s = 0.0105$) after which there is no turbulent burst and the flow starts to laminarize after an initial stable period. If slip length is further increased, the initial stable period starts to reduce in duration and eventually the flow immediately laminarizes with negligible initial stable period for very large slip lengths. These observations for the P4 solution suggest that a slip surface appears to stabilize a flow and promote a return to the stable laminar state beyond the critical value ($L_s = 0.0105)$. Thus, it is evident that the slip surfaces provide distinct effects on the laminar-turbulent separatrix of the P3 and P4 solutions.

\begin{figure}
 \begin{center}\includegraphics[trim=2 2 2 2,clip,scale=.625]{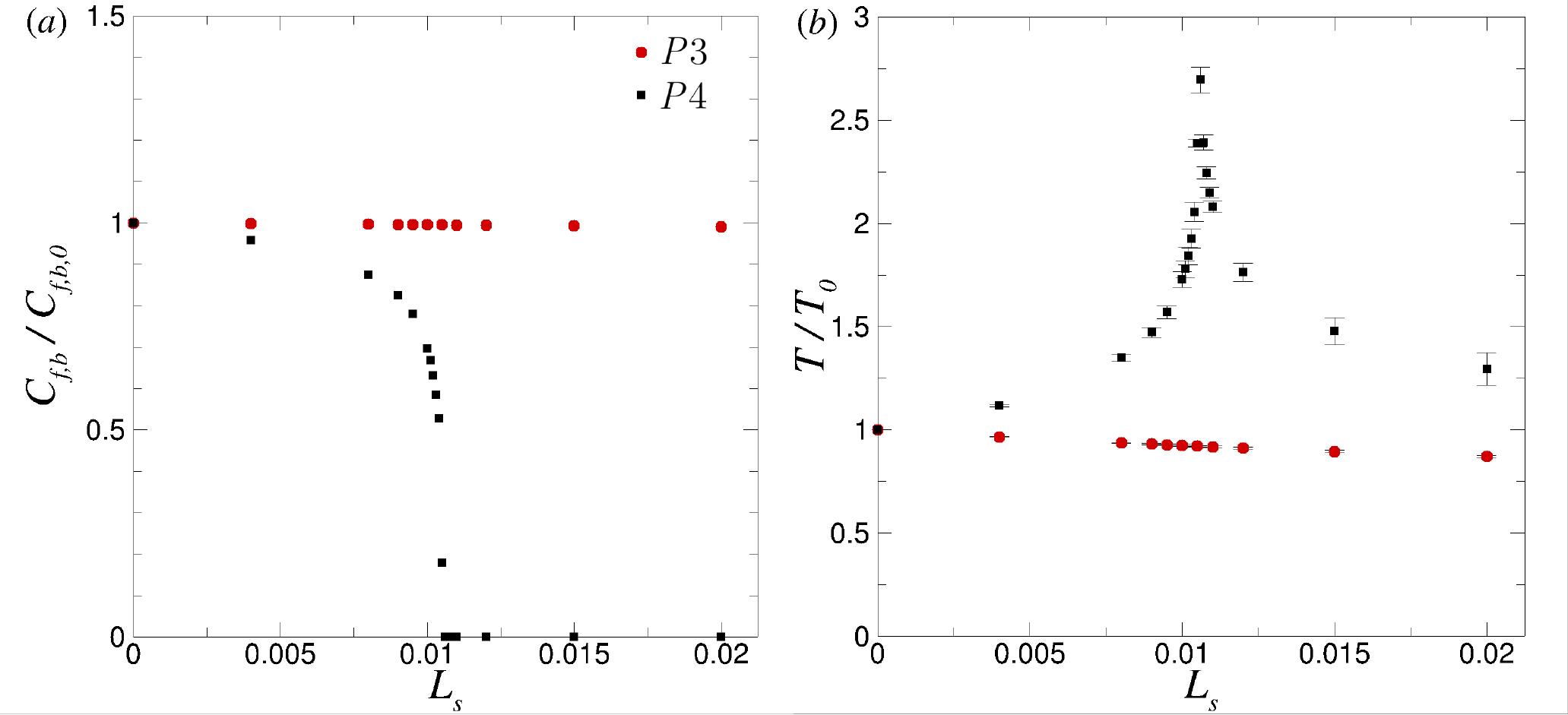}
  \caption{(\textit{a}) Maximum skin-friction coefficient $C_f$ in turbulent burst and (\textit{b}) duration of initial stable period. The duration of stable period is defined by the time for which $C_f$ is between $\pm10\%$ of $C_{f,0}$ and normalized by the duration of the initial stable period for the no-slip case. Lower and upper error bars correspond to values when using $\pm5\%$ and $\pm15\%$ of $C_{f,0}$, respectively.
    \label{fig:stableburst}}
 \end{center}
\end{figure}

To further characterize the temporal dynamics observed in the behaviors of the skin friction, figures \ref{fig:stableburst}$(a)$ and $(b)$ show the maximum skin-friction achieved during the turbulent bursting period and the duration of the initial stable period, respectively. Here, we define the bursting skin friction coefficient as $C_{f,b} = C_{f,max} - C_{f,0}$, where $C_{f,max}$ is the maximum skin-friction at the end of the bursting period and $C_{f,0}$ is skin friction at initial time. The stable period, $T$, is defined as the duration for which skin friction continues to stay between $\pm10\%$ of the initial skin-friction value depending on the direction of the trajectories: 10\% for bursting direction and -10\% for laminarization direction. Both quantities are normalized by the associated values of the no-slip case. For the P3 solution, the magnitude of the turbulent burst remains relatively constant with slip length -- the slip surface has a negligible effect on skin friction reduction during the strong burst phase of the transition. The stable period slightly decreases with slip length and for a very substantial slip length of $L_s = 0.02$, the stable period is reduced by approximately 10\% of the no-slip case. This observation indicates that slip surfaces cause the turbulent burst to occur sooner with almost same magnitude, promoting an early transition to turbulence. For the P4 solution, on the other hand, the magnitude of the turbulent burst decreases almost asymptotically with slip length up to a critical value of $L_s = 0.0105$. After the critical slip length, there are no turbulent burst events because the flow becomes laminarized. Interestingly, the stable period of the P4 solution shows a non-trivial behaviour. It increases until reaching a critical value at $L_s = 0.0105$ and starts to decrease because laminarization occurs sooner with slip length. Before the critical slip length, the slip length appears to play a role in delaying the transition to turbulence, as opposed to the P3 solution. It is worth noting that the dependence of the Reynolds number was tested by using different values of the Reynolds number in the transitional regime, where almost identical trends were produced for the P3 and P4 solutions (not shown).

\subsubsection{Linear growth rate in transition}

We estimate the growth rate of the linearly stable stage along the turbulent burst trajectories for the P3 and P4 solutions by introducing a time-dependent variable $s(t) = \vert\vert \boldsymbol{u}(t) \vert\vert - \vert\vert \boldsymbol{u}(t_0) \vert\vert$, where $\vert\vert  \boldsymbol{u}(t) \vert\vert$ and $\vert\vert \boldsymbol{u}(t_0) \vert\vert$ are the $L^2$-norm of the velocity field at time $t$ and initial time $t_0$, respectively. Given the definition of the turbulent burst as an escape from the exact coherent solutions consisting of linearly unstable and nonlinear unstable stages, the linear growth rate can be estimated by fitting the linear portion of a bursting trajectory observed in $s(t)$ to an exponential function $Ae^{\sigma t}$, where $A$ is a constant related to the magnitude of the unstable eigenvector and $\sigma$ is the associated growth rate \citep{Gibson2008}.

\begin{figure}
 \begin{center}\includegraphics[trim=2 2 2 .5,clip,scale=.625]{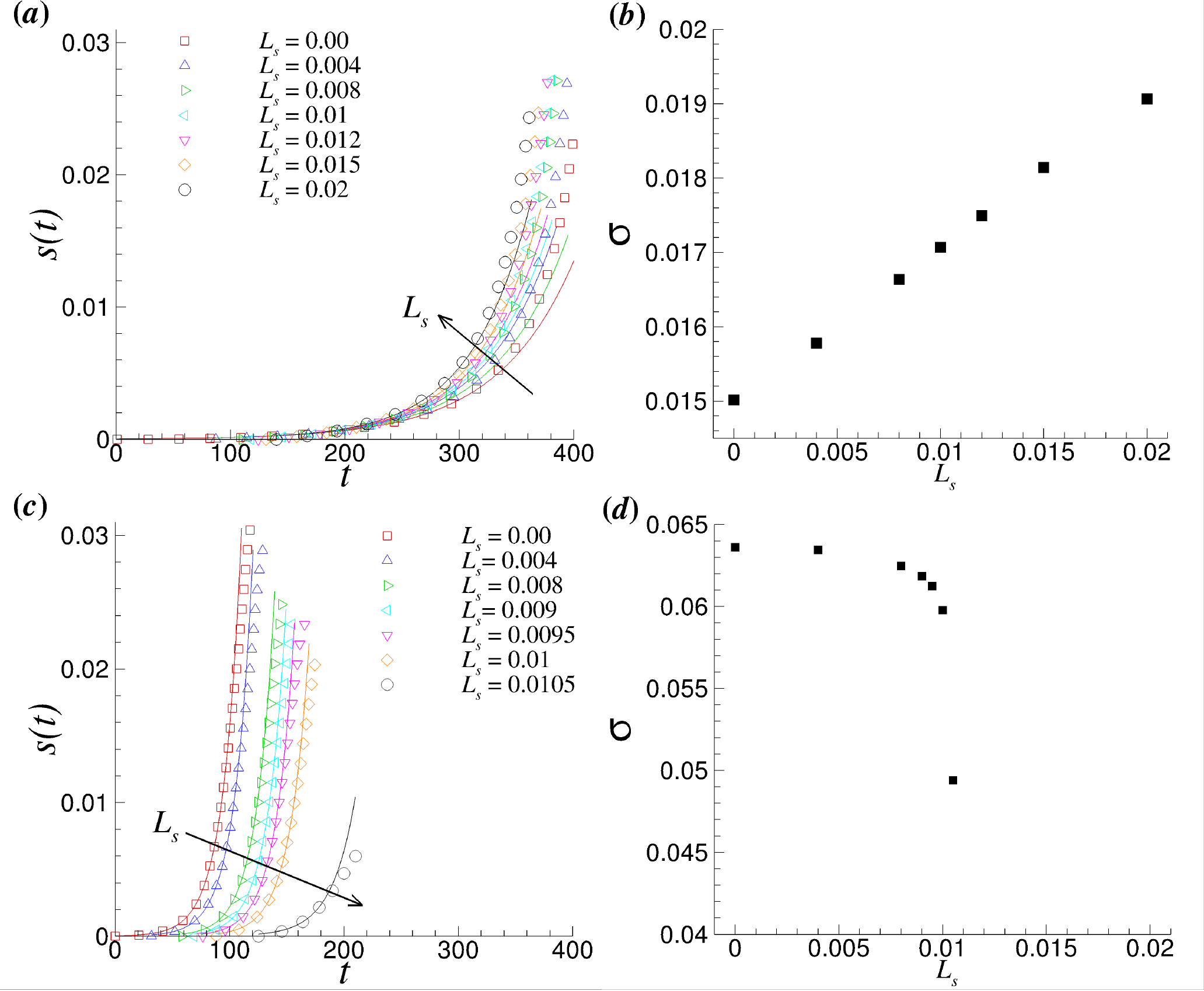}
  \caption{Growth rates for the P3 ($a,b$) and P4 ($c,d$) solutions for various slip lengths. The growth rate can be approximated for short time during which the system behaves like exp($\sigma t$), where $\sigma$ s the linear growth rate.
    \label{fig:growth}}
 \end{center}
\end{figure}

Figure \ref{fig:growth} shows the evolution of $s(t)$ and the growth rate of the P3 and P4 solutions for various slip lengths. The growth rate for the P3 solution tends to increase linearly with slip length, which is not all surprising given the trend seen in the skin friction in figure \ref{fig:Cf}($a$). Conversely, the growth rate for the P4 solution decreases almost asymptotically with slip length. This same trend was observed in the turbulent burst value in figure \ref{fig:stableburst}($a$). Again, the escape from the P3 solution occurs earlier with increasing slip length, while the escape is delayed for the P4 solution. It should also be noted that the values of the growth rate for the P3 solution are lower than those for the P4 solution as the P3 is believed to be closer to laminar state \citep{Park2015jfm}. In addition, since the escape process from an exact coherent solution follows closely its unstable manifold associated with the most unstable eigenvalue, those growth rate values are comparable to the most unstable eigenvalues of these solutions \citep{Park2015jfm,Park2018}.

\subsubsection{Phase-space dynamics with slip surfaces}
Here, we investigate the phase-space dynamics for the P3 and P4 solutions over slip surfaces by projecting the dynamics onto the $I$-$D$ space, where $I$ and $D$ are the energy input rate and energy dissipation rate, respectively. For the current study (Poiseuille flow), the energy input rate is given as
\begin{equation} \label{eq:input}
I=\frac{1}{2 L_z}\int_{0}^{L_z}\int_{-1}^{1}(p u\vert_{x=0} - p u\vert_{x=L_x})dydz
\end{equation}
\\
\noindent and the energy dissipation rate is given as
\begin{equation} \label{eq:dissipation}
D=\frac{1}{2 L_x L_z}\int_{0}^{L_z}\int_{-1}^{1}\int_{0}^{L_x}(\vert\nabla u \vert^2 + \vert\nabla v \vert^2 + \vert\nabla w \vert^2)dxdydz.
\end{equation}
\\
%\noindent for $j = 1,2,3$ which correspond to streamwise, wall-normal, and spanwise directions, respectively. 
The total energy of the flow is defined by
\begin{equation}
E = \frac{1}{2 L_x L_z}\int_{0}^{L_z}\int_{-1}^{1}\int_{0}^{L_x}(u^2 + v^2 + w^2)dxdydz
\end{equation}  
and, thus, the rate of change in energy for the flow is equal to $dE/dt = I - D$. For exact coherent solutions, $D = I$.

\begin{figure}
	\includegraphics[trim=2 2 2 2,clip,scale=.65]{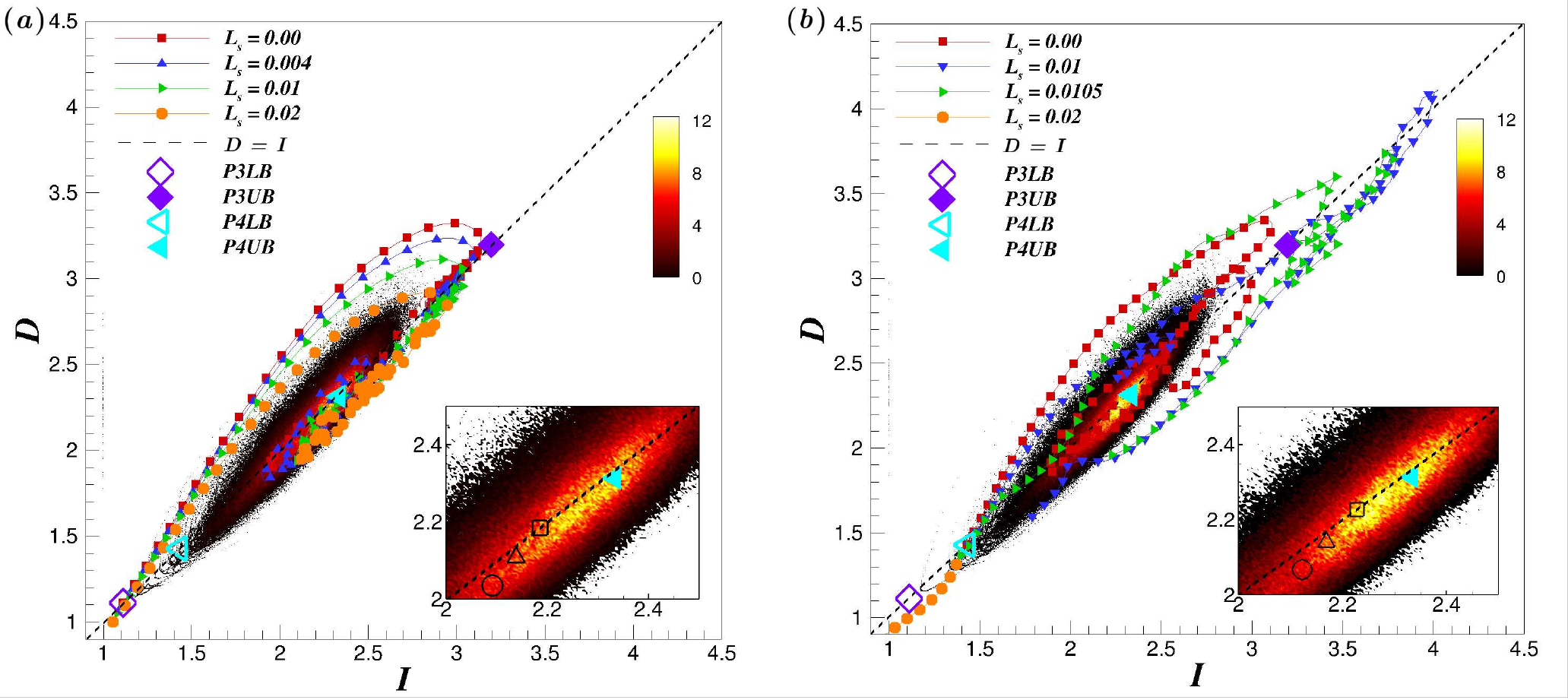}
  \caption{The phase space projection of turbulent trajectories onto the energy input rate ($I$) and energy dissipation rate ($D$) for various $L_s$ at ($a$) $Re_c$ = 1760 and ($b$) $Re_c$ = 1800. The black dashed line shows an equiliburm line ($D=I$). P3 lower (\Diamondshape) and upper (\FilledDiamondshape) and P4 lower (\TriangleLeft) and upper (\FilledTriangleLeft) solutions are also shown \citep{Park2015jfm}. The contour shows the probability density function (PDF) of the $I-D$ state for the no-slip case at ($a$) $Re_c = 1760$ and ($b$) $Re_c = 1800$. Inset: The mean states of long-time flows at each respective $Re_c$ for $L_s = 0.00$ (\Square), $L_s = 0.01$ (\TriangleUp), and $L_s = 0.02$ (\Circle) overlaid on the no-slip PDF.} 
    \label{fig:IDplane}
\end{figure}

Figures \ref{fig:IDplane}($a$) and ($b$) shows the $I$-$D$ phase diagrams of transition and turbulent trajectories starting from P3 and P4 lower-branch solutions, respectively, along with their corresponding upper-branch solutions. A joint probability density function (PDF) of turbulent trajectories for no-slip case is also plotted. In figure \ref{fig:IDplane}($a$), all trajectories from the P3 lower-branch solution with different slip lengths exhibit a qualitatively similar trend. Starting from the lower-branch solution, they approach the upper-branch solution, make a turn around the equilibrium line ($D=I$), and move toward a core region of PDF. However, as slip length is increased, the trajectory tends to approach the equilibrium line earlier and then move toward the core region. Therefore, larger slip lengths allow a flow to approach a turbulent state faster, promoting an earlier transition to turbulence, as seen in figure \ref{fig:Cf}($a$). Figure \ref{fig:IDplane}($b$) shows the trajectories starting from the P4 lower-branch solution, exhibiting similar trends to those seen in the P3 solution up to the critical slip length $L_s = 0.0105$ -- the trajectory gets closer to the equilibrium line as slip length is increased. Around the critical slip length, however, trajectories are altered drastically, bypassing the P4 upper-branch solution and approaching the P3 upper-branch solution before laminarizing. Beyond $L_s = 0.0105$, the trajectories immediately approach the laminar state with no bursting trajectories, as seen in figure \ref{fig:Cf}($b$).

It is evident that slip surfaces lead to modifications of the turbulent trajectories in the $I$-$D$ phase-space. The core region of the turbulent trajectories with slip surfaces deviates from that of the no-slip case -- it gets closer to the lower-branch solutions. These observations suggest that the distance between the turbulent state and the laminar-turbulent separatrix is reduced with the inclusion of slip surfaces. This is illustrated by the insets in figure \ref{fig:IDplane}($a$) \& ($b$) showing the mean state of the system for $L_s = 0.00, 0.01,$ and $0.02$. As slip length is increased, the mean state of the system shifts closer to the lower-branch solutions. This reduced distance indicates a greater likelihood for the flow to approach the separatrix and, thus, a greater likelihood that the flow will laminarize. Similar behavior has also been observed by applying opposition control schemes to wall-bounded turbulent flows \citep{Ibrahim2019b}.

\subsubsection{Flow structures}\label{subsec:structures}
\begin{figure}
 \begin{center}\includegraphics[trim=2 2 4 1,clip, scale=.65]{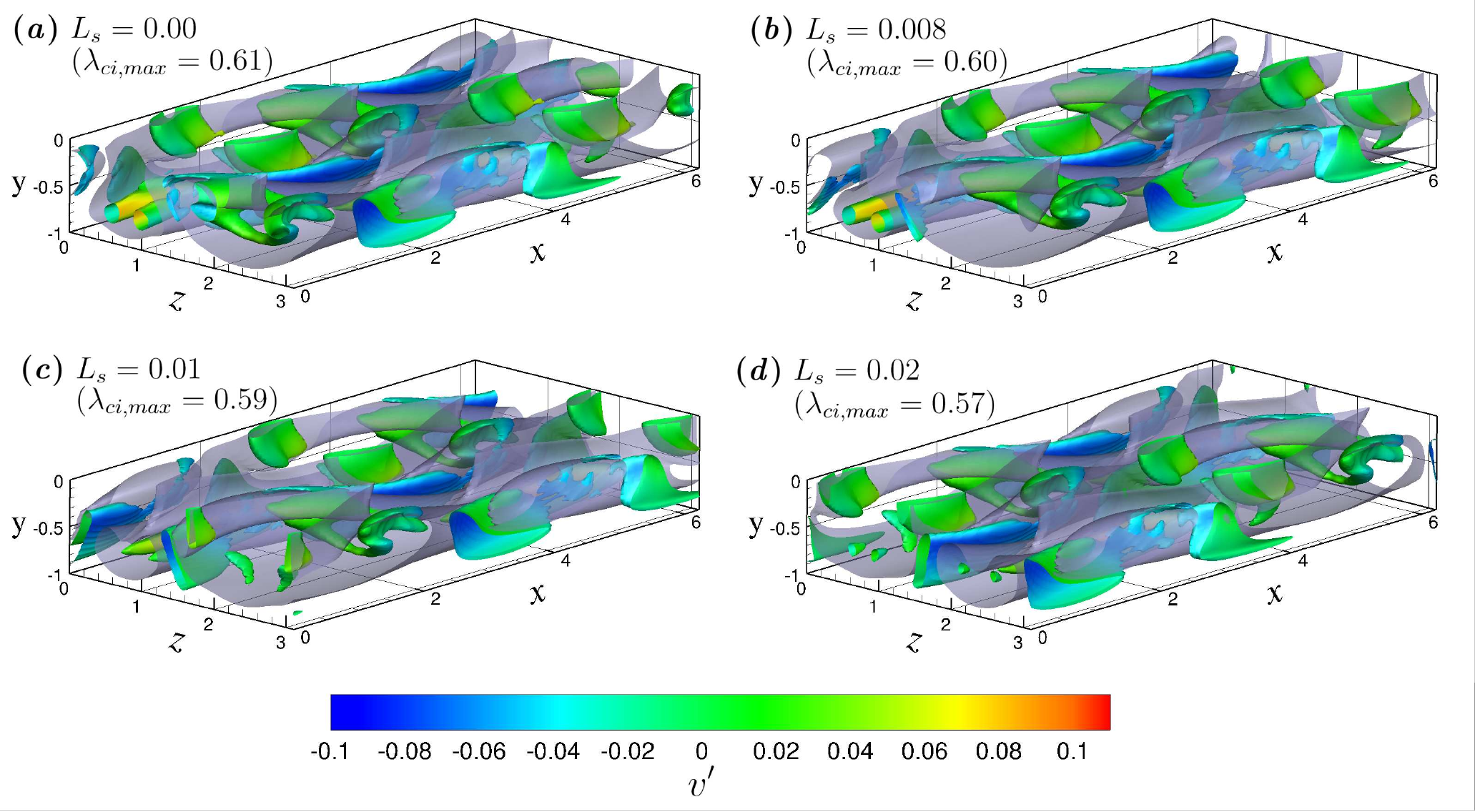}
  \caption{Vortical structures of the P3 lower branch solution at $Re_c = 1760$ for slip lengths $(a)$ $L_s = 0.00$, $(b)$ $L_s = 0.008$, $(c)$ $L_s = 0.01$, and $(d)$ $L_s = 0.02$. The maximum swirling strength for each case is presented in parentheses. The multicolored tubes are isosurfaces of 1/2 of maximum swirling strength. The contours flooding the isosurfaces represent the wall-normal velocity. Minimum and maximum values of wall-normal velocity represented by the contours correspond to -0.1 (blue) and 0.1 (red), respectively. The gray isosurface represents the critical layer, where local streamwise velocity is equal to the wave speed of the traveling wave solution.
    \label{fig:P3structures}}
 \end{center}
\end{figure}

\begin{figure}
 \begin{center}
 \includegraphics[trim=2 2 4 1,clip, scale=.65]{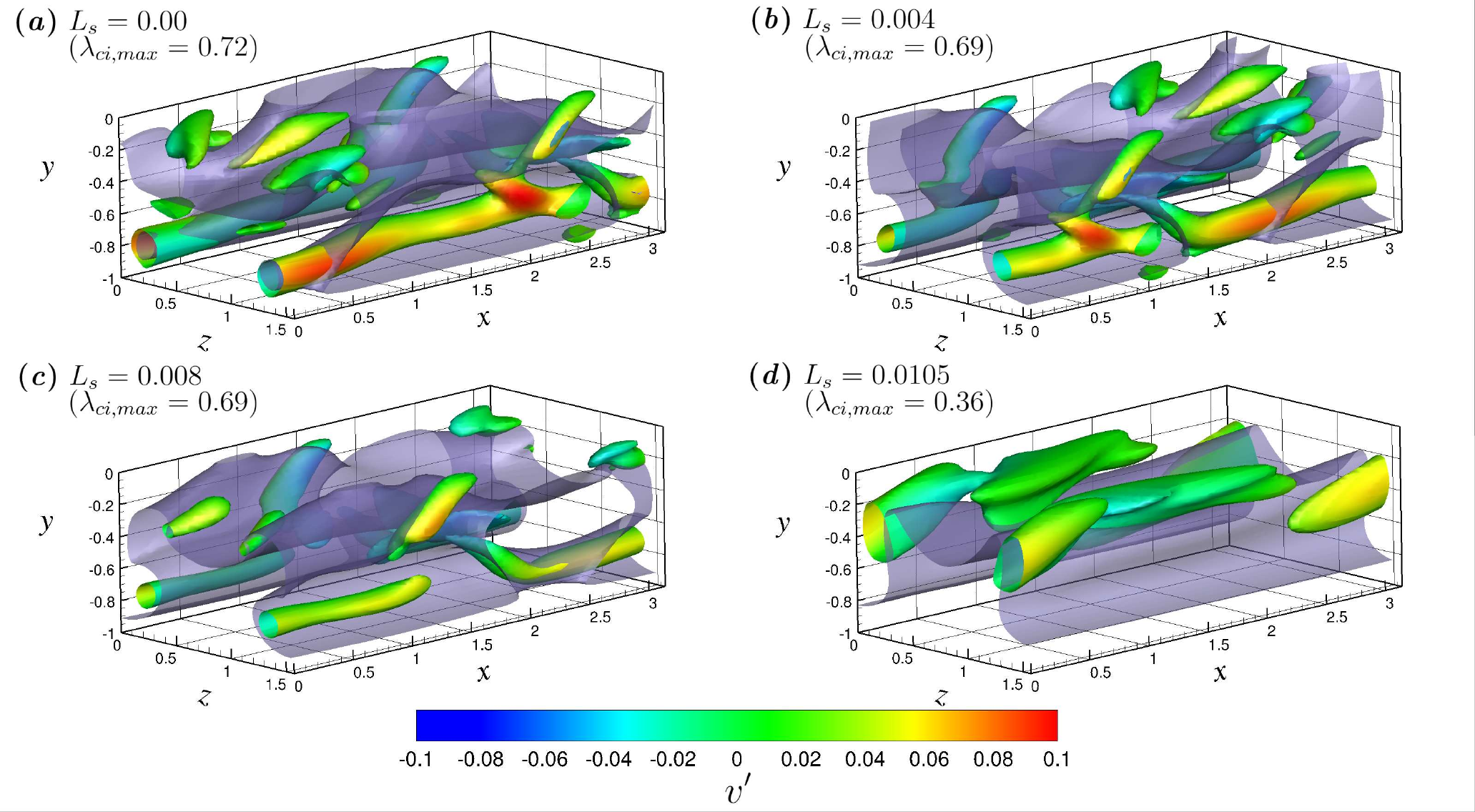}
  \caption{Vortical structures of the P4 lower branch solution at $Re_c = 1800$ for slip lengths $(a)$ $L_s = 0.00$, $(b)$ $L_s = 0.004$, $(c)$ $L_s = 0.008$, and $(d)$ $L_s = 0.0105$. The maximum swirling strength for each case is presented in parentheses. The multicolored tubes are isosurfaces of 1/2 of maximum swirling strength. The contours flooding the isosurfaces represent the wall-normal velocity. Minimum and maximum values of wall-normal velocity represented by the contours correspond to -0.1 and 0.1, respectively. The gray isosurface represents the critical layer, where local streamwise velocity is equal to the wave speed of the traveling wave solution.
    \label{fig:P4structures}}
 \end{center}
\end{figure}

We investigate the effects of slip surfaces on flow structures for which the swirling strength $\lambda_{ci}$, the imaginary part of the complex conjugate eigenvalues of the velocity gradient tensor \citep{ZhouJFM1999}, are calculated. Figures \ref{fig:P3structures} and \ref{fig:P4structures} show contours of the swirling strength for the lower half of the channel for the P3 and P4 solutions, respectively, at the end of the turbulent bursting trajectory. The contours represent isosurfaces of $50\%$ of the maximum swirling strength for each solution, which is also given in the figure. The wall-normal velocity is represented by the colour contours flooding the isosurfaces. These colour contours provide insights into the effect of slip surface on the vertical motion of these vortical structures.

As seen in \cite{Park2015jfm}, the P3 lower-branch solution displays large vortex cores near the channel centre -- this solution is called core mode. As the bursting process proceeds, the large vortices appear to be broken into smaller ones, move toward the wall, and spread across the span of the domain. Eventually, there is a large population of small vortex cores across the channel, as shown in figure \ref{fig:P3structures}. 
It appears that the general shape and position of the structures is mostly unchanged for all slip lengths studied, while the maximum swirling strength is slightly reduced with slip length. The majority of the wall-normal velocities associated with these vortex structures are close to zero at this instant, while a very strong negative wall-normal velocity is located around vortex structures during bursting trajectories.

In figure \ref{fig:P4structures}, the vortical structures of the P4 solution are presented. As shown by \cite{Park2015jfm}, the P4 lower-branch solution displays vortices of $\Lambda$-like structures without a head connecting the legs. The vortex cores are located near the critical layer where the local streamwise velocity matches the wave speed -- this solution is called critical-layer mode. Recently, this solution family was continued at higher Reynolds numbers, showing hairpin-like vortex structures \citep{Shekar2018}. With slip surfaces, there still appear to be leg structures with slightly upward side branches, which is also observed in a trajectory along the most unstable manifold of the P4 lower-branch solution \citep{Park2018}. Its structures are significantly longer than ones of the P3 solution and are accompanied by a small number of smaller vortex cores around these long vortex structures. Vortex cores of the P4 solution seem to not extend into the channel centre but remain mostly streamwise-oriented. As slip length is increased, the maximum swirling strength is drastically reduced (i.e. $~50\%$ at $L_s = 0.0105$). In particular, from the colour contours of the wall-normal velocity associated with long vortex structures, the vortex structures are shifted upward from the wall as slip length is increased, leaving a quiescent region near the wall.
 
\begin{figure}
\begin{center}
\includegraphics[trim = 0 2 1 0, clip, scale=.6]{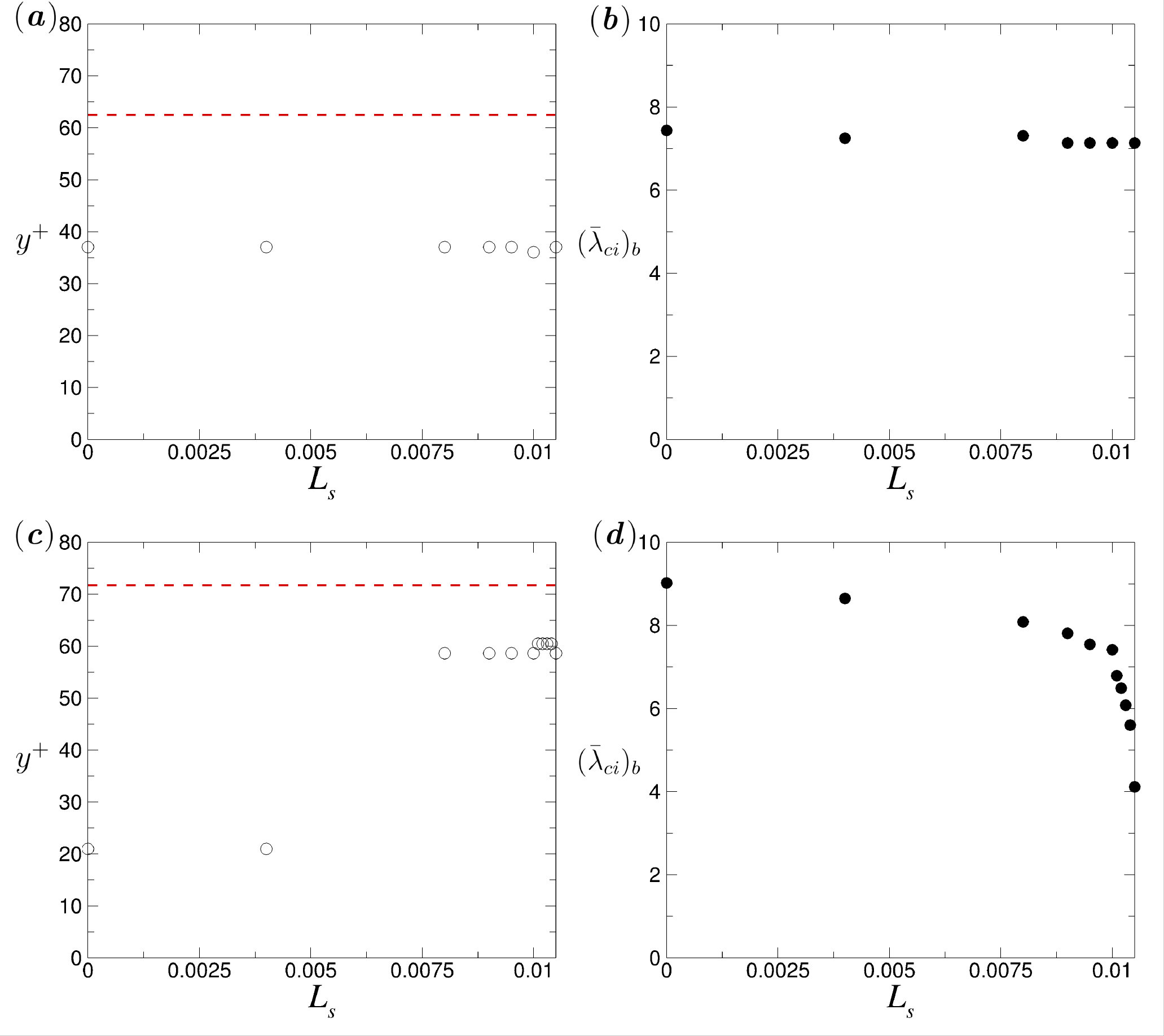}
\caption{($a,c$) The wall-normal location corresponding to the maximum of the area-averaged swirling strength for the P3 and P4 solutions, respectively. The dashed line corresponds to the channel centre for the P3 and P4 solutions. ($b,d$) The bulk swirling strength as a function of slip length for the P3 and P4 solutions, respectively. All values are calculated at the time at which the maximum $L^2$-norm of velocity is reached.
\label{fig:lambdaci}}
\end{center}
\end{figure}

To quantitatively investigate the effect of the slip surfaces on the vertical shifting of vortical structures, the wall-normal location for the maximum value of the area-averaged swirling strength is plotted in figures \ref{fig:lambdaci}($a$) and \ref{fig:lambdaci}($c$) for P3 and P4 solutions, respectively. Bulk swirling strength is also shown for the P3 and P4 solutions in figure \ref{fig:lambdaci}($b$) and \ref{fig:lambdaci}($d$), respectively. The wall-normal location of the maximum of the average swirling strength for the P3 solution is almost constant for all slip lengths studied, as shown in figure \ref{fig:lambdaci}($a$). Figure \ref{fig:lambdaci}($b$) also shows that the effect of slip length on the bulk swirling strength is minimal, suggesting the slip surfaces have a minimal effect on the vortex structures associated with the P3 solution. However, vortex structure dynamics for the P4 solution are quite different than those of the P3 solution. There are much larger deviations in height of maximum swirling strength as slip length is increased. Beyond $L_s=0.004$, the vortex structures are significantly shifted upward away from the wall, moving toward the channel centre. The bulk swirling strength is also significantly decreased as slip length increases as seen in figure \ref{fig:lambdaci}($d$). The trend appears to be asymptotic with slip length. Slip surfaces seem to have a profound weakening effect on the P4 vortex structures. These observations for the P4 solution are in good agreement with the experimental study \citep{Zhang2015g}, where the mechanisms for drag reduction over superhydrophobic surfaces were investigated for a turbulent boundary layer flow. They performed experiments to show that the drag reduction is caused by a combination of slip at the surface and modifications to the turbulent structures -- vortices are weakened and lifted away from the surfaces. The combination of these modifications to the vortical structures (i.e. weakened and shifted vortices) of the P4 solution help to explain why, beyond the critical slip length, there is no turbulent burst and flow laminarizes. It appears that slip surfaces modify the turbulent structures such that the self-sustaining cycle of near-wall turbulence is disrupted and turbulence can no longer be sustained.

\subsubsection{Spatiotemporal dynamics}\label{subsec:spatiotemporal}
We now attempt to illuminate the mechanisms behind early or delayed transition to turbulence observed in the P3 and P4 solutions. It has been shown that at the onset of turbulence, the spatiotemporal dynamics provides a coherent mechanistic basis for the dynamics of transition to turbulence \citep{Avila2011,Shih2016,Barkley2015} -- puffs and slugs or directed percolation have been explored in this regard. Similarly, the spatiotemporal dynamics of the exact coherent solutions also appear to provide a coherent mechanistic basis for effects of slip surfaces on the transition observed in the P3 and P4 solutions.

For spatiotemporal dynamics, we incorporate a Fourier decomposition of the velocity field. Similar with \cite{Wang2007}, the velocity field of the traveling wave solutions can be Fourier decomposed in the $x$ direction as
\begin{equation}
\boldsymbol{u}(x,y,z,t) = \boldsymbol{u}_0(y,z,t) + \sum_{n=1}^{\infty} \left( \textrm{e}^{i n \theta} \boldsymbol{u}_n(y,z) + \textrm{c.c.} \right),
\end{equation}
where $\alpha$ is the fundamental wave number in the streamwise direction, $\theta = \alpha(x-ct)$, $c$ is the constant wave speed, and c.c. denotes complex conjugates. For this study, we define the 0-mode as $\boldsymbol{u}_0(y, z, t) = (u_0, v_0, w_0)$. The streamwise fluctuation $u'$ can then be calculated by $u_0(y, z, t) - \overline{u}(y)$, where $\overline{u}(y)$ is the time- and area-averaged mean velocity. Note that the wall-normal velocity fluctuation $v'(y, z, t) = v_0(y, z, t)$ because $\overline{v}(y)=0$. These 0-mode velocity fluctuations $u'$ and $v'$ provide information about streaky flow and rolls, respectively \citep{Wang2007}. Although $u'$ and $v'$ represent streamwise-averaged fluctuations, these quantities still enable identifying the sweep and ejection events based on the quadrant analysis \citep{Picella2019}.

\begin{figure}
\begin{center}
\includegraphics[angle=90,origin=c,scale=.85, trim=1 2 2 1, clip]{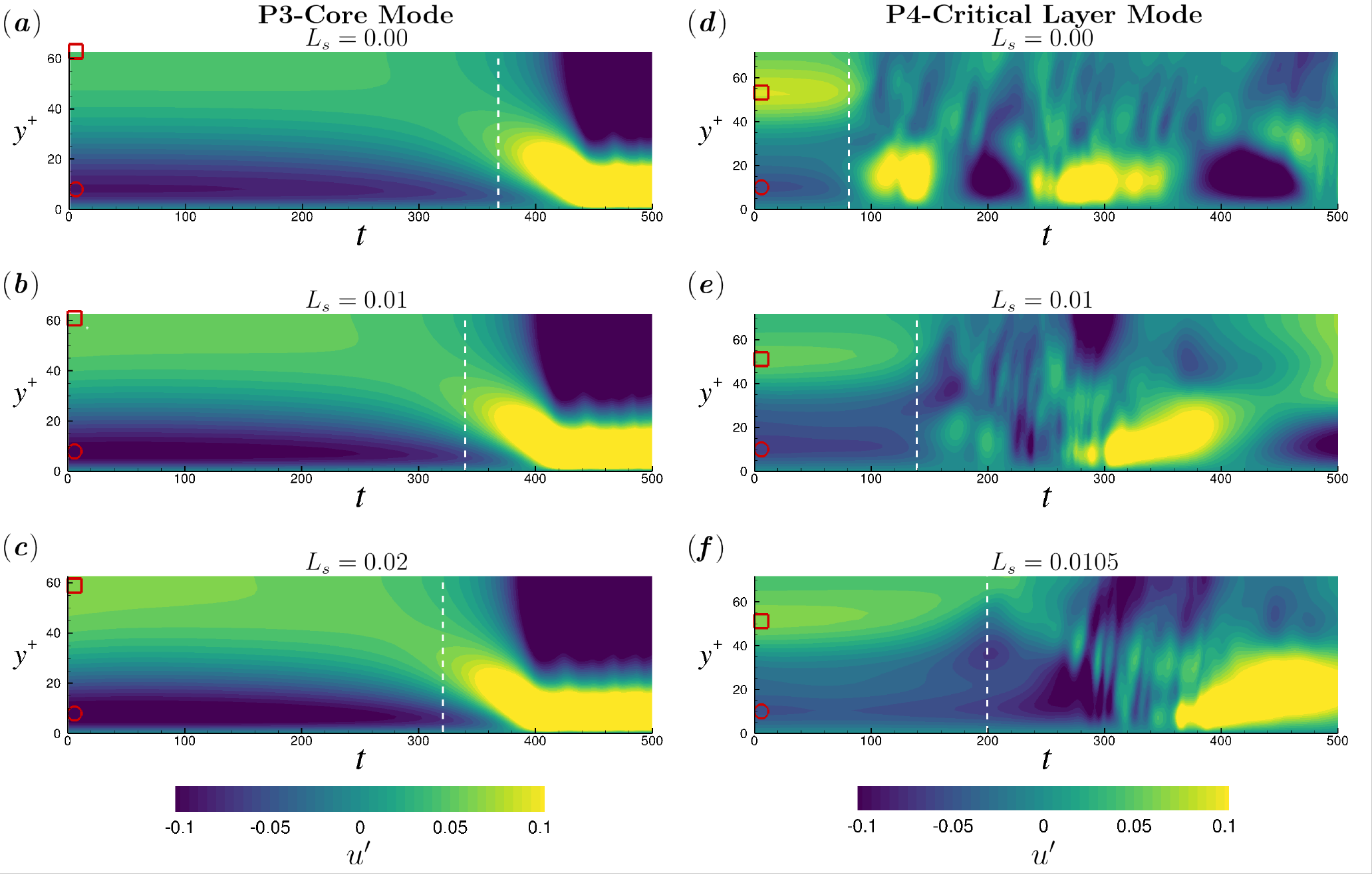}
\caption{The streamwise-averaged streamwise velocity fluctuations $u'$ at  $z^+ = L_z^+/4$ as a function of wall-normal distance for the P3 solution $(a)$-$(c)$ and the P4 solution $(d)$-$(f)$ for various slip lengths. The dashed white line corresponds to time of transition ($C_f = 1.1C_{f,0}$). The symbols (\Circle) and (\Square) are indicated for the minimum and maximum $u'$ at their wall-normal locations, respectively. For the P3 solution, ($u'_{min}, u'_{max}$) = (-0.0981, 0.0461) for ($a$), (-0.1110, 0.0569) for ($b$),  (-0.1223, 0.0643) for ($c$). For the P4 solution, ($u'_{min}, u'_{max}$) = (-0.0665, 0.1259) for ($d$), (-0.0248, 0.0917) for ($e$), (-0.0262, 0.1008) for ($f$). Note that these values remain almost constant before the time of transition.
\label{fig:spatiotemporal_uf}}
\end{center}
\end{figure}

\begin{figure}
\begin{center}
\includegraphics[angle=90,origin=c,scale=.85, trim=1 2 2 1, clip]{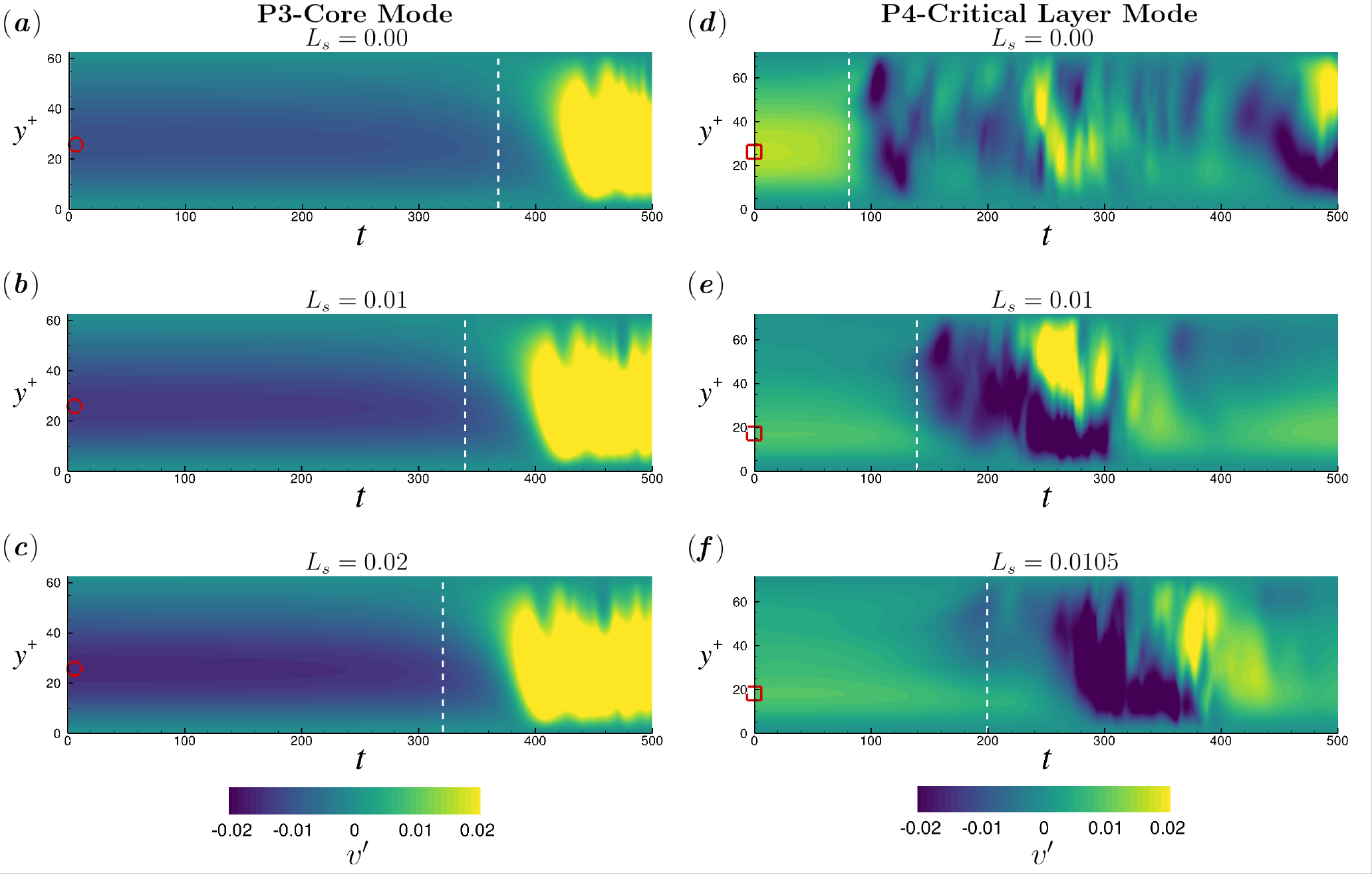}
\caption{The streamwise-averaged wall-normal velocity fluctuations $v'$ at $z^+ = L_z^+/4$ as a function of wall-normal distance for the P3 solution $(a)$-$(c)$ and the P4 solution $(d)$-$(f)$ for various slip lengths. The dashed white line corresponds to time of transition ($C_f = 1.1C_{f,0}$). The symbols (\Circle) and (\Square) are indicated for the minimum and maximum $v'$ at their wall-normal locations, respectively. For the P3 solution, ($a$) $v'_{min} = -0.0107$, ($b$) $v'_{min} = -0.0139$, ($c$) $v'_{min} = -0.0158$. For the P4 solution, ($d$) $v'_{max} = 0.0181$, ($e$) $v'_{max} = 0.0131$, ($f$) $v'_{max} = 0.0147$. Note that these values remain almost constant before the time of transition.
\label{fig:spatiotemporal_vf}}
\end{center}
\end{figure}

Figures \ref{fig:spatiotemporal_uf} and \ref{fig:spatiotemporal_vf} show the spatiotemporal dynamics of streamwise and wall-normal velocity fluctuations at a spanwise location ($z^+=L_z^+/4$) for different slip lengths, respectively, along with the dashed line corresponding to time of transition. This spanwise location is chosen because it is the location where the core of vortex structures is located -- the other location ($z^+=3L_z^+/4$) where the core of vortex structures is also located was studied, giving essentially identical results.

Both solutions exhibit similar behaviours on the streamwise fluctuations. For the P3 solution, as shown in figures \ref{fig:spatiotemporal_uf}$(a)$-$(c)$, the near-wall low-speed streak (negative $u'$) appears to be enhanced as slip length is increased ($u'_{min}$ decreases from -0.0981 to -0.1223), while for the P4 solutions, the near-wall low-speed streak is slightly extended upward but remains relatively quiescent ($u'_{min} \approx -0.025$ with slip surfaces) compared to the P3, as seen in figures \ref{fig:spatiotemporal_uf}$(d)$-$(f)$. Near the channel centre, there are positive streamwise fluctuations for both solutions. In particular, a height of peak streamwise fluctuations for the P3 solution corresponds to the region where the core of vortex structures is localized.

Most interestingly, the wall-normal fluctuations show the opposite behaviours. In figures \ref{fig:spatiotemporal_vf}$(a)$-$(c)$, the wall-normal fluctuations of the P3 solution are all negative across the channel height and become more negative in the buffer layer with increasing slip length ($v'_{min}$ decreases from -0.0107 to -0.0158). The combination of increased streamwise velocity near the channel centre and strong negative wall-normal velocity in the buffer layer seems to promote an instability to the vortex structures at the channel centre. Specifically, this instability causes the large vortex structures to propagate toward the wall much sooner where they break up as the slip length is increased. These effects seem to lead to an early transition for the P3 solution. On the other hand, for the P4 solution, figures \ref{fig:spatiotemporal_vf}$(d)$-$(f)$ show all positive wall-normal fluctuations across the channel height and peak values are located around $y^+ = 20-40$, where the core of the vortex structures is localized. As slip length increases, the peak value decreases as opposed to the P3 solution (from $v'_{max} =$ 0.0181 to 0.0147). The combination of the very quiescent flow and positive wall-normal velocity around the vortex structures appears to allow the the vortex structures to be more stable and persist for a longer time before transition to turbulence occurs. Those effects appear to delay the transition. It is interesting to note that based on the quadrant analysis, slip surfaces are likely to promote stronger wall-toward motions (Q4-like) at the channel centre and inward interactions (Q3) in the buffer layer for the P3 solution, while for the P4 solution, the surfaces tend to promote longer ejection events (Q2) near the critical layer and wall-away motion (Q1-like) at the channel centre. This quadrant analysis may help to elucidate distinct transition dynamics for the P3 and P4 solutions -- the combination of the core mode and strong sweep events seems to promote early transition, while the combination of the critical layer mode and extended ejection events seems to delay the transition.

Finally, we can make the link between transition dynamics and flow structures, as the latter can be thought of as different disturbances in a receptivity process of transition \citep{Kachanov1994}. Flow structures associated with the core mode (P3 solution) seem to resemble those found in free-stream turbulence (i.e., more densely located at the centre of the channel or the edge of the boundary layer) as seen in figure \ref{fig:P3structures} -- it can be called external perturbations.  On the other hand, the P4 solution (critical layer mode) exhibits $\Lambda$-shaped vortical structures, confined near the wall, resembling three-dimensional flow structures developed from Tollmien-Schlichting waves  -- it can be called internal perturbations. Depending on different disturbances, transition scenarios appear to be very different. \cite{Sayadi2013} compared the dynamics of different transition scenarios, namely, H-type, K-type, and bypass transition. In the H- and K-type transitions, characterized by $\Lambda$-shaped vortical structures, a large overshoot in the skin-friction coefficient was observed when transition starts to occur. However, for the bypass transition, this overshoot was absent, giving a smooth transition to turbulence. This distinct overshoot is present in the skin-friction coefficient profile for the P4 solution and absent in that for the P3 solution as shown in figure \ref{fig:Cf}. It is worth noting that, similar to the P3 solution, \cite{Wu2009} introduced free-stream turbulence to a boundary layer and showed that the bypass transition was triggered. They showed that the typical near-wall Tollmien-Schlichting waves were not the initiating mechanism for transition and in this case was attributed to a bypass route triggered from the free stream. The evolution of the P3 solution is similar in that the structures originate in the centre of the channel (i.e. free-stream) and propagate downward toward the near-wall region before breaking down, subsequently triggering transition. In addition, the shape of the vortical structures for the P4 solution are similar to those found in H- and K-type transitions. A similar observation for the effects of slip surfaces on the H- and K-type transition was also made by \citet{Picella2019}, where the overshoot in skin-friction was present and transition was delayed by the slip surface with the modal perturbation. However, they also showed that the non-modal perturbation, similar to those perturbations that cause bypass transition, was unaffected by the slip surfaces, contrary to the early transition behavior observed in the P3 solution of the current study. Differences in the transition behavior of the non-modal perturbation of \citet{Picella2019} and the P3 solution of the current study could stem from assumptions made in the linearization of the Navier-Stokes equations or, possibly, differences between the non-modal perturbation and the P3 solution. The differences in the skin-friction coefficient and vortical structures between the P3 and P4 solutions may provide clear and plausible mechanisms responsible for early transition for P3 and delayed transition for P4 due to slip surfaces, for which further investigation is yet needed.

\section{Conclusion}\label{sec:conclusion}
Direct numerical simulations were performed to investigate the effect of slip surfaces on transition in plane Poiseuille (channel) geometry. For the purpose of validations, laminar drag reduction values were calculated and compared to the previous studies. Levels of over 10\% drag reduction were observed and in good agreement with previous numerical and experimental studies. The drag reduction percentage remains almost constant regardless of the Reynolds number. In particular, our results are in great agreement with the theory for drag reduction of superhydrophobic surfaces on both walls in a laminar channel flow.

Turbulence lifetime analysis (i.e. the probability that turbulence will persist) was investigated for transitional flows at $\Rey_c = 1600, 1800,$ and $2000$ $(\Rey_\tau=77,$ $85,$ and $93$). Flows with slip surfaces were significantly less likely to maintain turbulence compared to the no-slip case. Additionally, the slip flows were more likely to laminarize at earlier times as slip length is further increased. As Reynolds number is increased, this trend still holds, while a larger slip length is needed to obtain the same likelihood of laminarization found at lower Reynolds numbers. Phase-space projection of transitional trajectories on the energy input and dissipation rates showed a reduced distance between the laminar and turbulent states, which helps explain the increased likelihood of laminarization due to slip surfaces. 

Exact coherent solutions, specifically nonlinear travelling wave solutions, to the Navier-Stokes equations were used to investigate the effects of slip surfaces on the laminar-turbulent separatrix. The P3 and P4 solution families were chosen as their lower-branch solutions are shown to lie on the basin boundary between laminar and turbulent flow \citep{Park2015jfm}. The skin-friction evolution and linear growth rate from the lower-branch solutions were calculated. For slip flows, the strong turbulent burst associated with the P3 lower-branch solution was induced at earlier times while the bursting magnitude and growth rate were mostly unaffected. For the P4 solution, however, the strong turbulent burst was delayed with the reduced magnitude and growth rate for slip flows. Beyond a critical slip value, the turbulent burst was completely eliminated as a flow is immediately laminarized after a short stable period. Effects of slip surfaces on vortex structures of the P3 and P4 solutions were examined to elucidate mechanisms responsible for the difference in transition behaviours between the solutions. It appears that structures associated with the P3 solution were largely unaffected by the slip surfaces. Overall structure and strength remained relatively constant. However, the strength of P4 vortex structures was weakened by $\sim50\%$, and they were shifted away from the wall. Based on the quadrant analysis and spatiotemporal dynamics, it was suggested that slip surfaces promote the prevalence of strong wall-toward motions (Q4-like events) in the area of the P3 vortex cores close to the channel centre. This results in instability, which promotes the propagation of the vortex structures down into the wall where they break up and induce transition, similar to the bypass transition \citep{Saric2002}. However, sustained ejection events (Q2) were present in the region of the P4 vortex cores (which resemble the $\Lambda$-shaped structures in H- and K-type transitions) resulting in a shift of the vortex structures away from the wall allowing them to remain intact and propagate downstream for a longer time. It can be suggested that the slip surfaces tend to affect core-mode structures (P3 solution) or non-modal perturbations via inward interactions (Q3) near the wall and wall-toward motions (Q4-like events) near the channel centre, subsequently leading to bypass-type early transition. On the other hand, the slip surfaces tend to affect critical-layer structures (P4 solution) or modal perturbations via ejection events (Q2) near the wall and wall-away motions (Q1-like events) near the channel centre, subsequently leading to H- and K-type delayed transition. These distinct transition dynamics of the P3 and P4 solutions due to slip surfaces could suggest that different flow control techniques could be used to delay or promote a transition to turbulence, which will be a subject of interesting future work.

\section*{Acknowledgements}
The authors gratefully acknowledge the financial support from the National Science Foundation through a grant OIA-1832976 and the Collaboration Initiative at the University of Nebraska. The authors are also grateful to George Gogos (University of Nebraska-Lincoln) for initiating this problem. The direct numerical simulation code used was developed and distributed by John Gibson at the University of New Hampshire. The authors also acknowledge the computing facilities used at the Holland Computing Center at the University of Nebraska-Lincoln.

\bibliographystyle{jfm}
% Note the spaces between the initials
\bibliography{refs}

\end{document}